\documentclass[journal]{IEEEtran}
\usepackage{verbatim}
\usepackage[pdftex]{graphicx}
\usepackage{amsmath,amssymb,amsfonts,bm}
\usepackage{epsf}
\usepackage{tikz}
\usepackage{psfrag}
\usepackage{pstool}
\usepackage{epstopdf}
\usepackage{stfloats}

\usepackage{url}

\usepackage[dvips]{epsfig}
\usepackage{amsthm}
\usepackage{color}
\usepackage[usenames,dvipsnames]{pstricks}
\usepackage{lettrine}
\usepackage[dvips]{epsfig}
\usepackage{pst-grad} 
\usepackage{pst-plot} 
\usepackage{epsfig}
\usepackage{enumerate}
\usepackage{amsbsy}
\usepackage{amssymb}
\usepackage{amsthm}
\usepackage{amscd}
\usepackage{float}
\usepackage[caption = false]{subfig}
\usepackage{color}
\usepackage[numbers,sort&compress,square]{natbib}

\usepackage{listings}
\usepackage{stackrel}
\usepackage{authblk}
\usepackage{dsfont}
\usepackage{algorithm}
\usepackage[noend]{algpseudocode}
\makeatletter
\newcommand{\removelatexerror}{\let\@latex@error\@gobble}
\makeatother
\newtheorem{thm}{Theorem}

\begin{document}
\title{Mobility-assisted Over-the-Air Computation for Backscatter Sensor Networks}

\author{Amin Farajzadeh, \and Ozgur Ercetin, and \and Halim Yanikomeroglu
\thanks{This work was supported by the European Unions Horizon 2020 Research
and Innovation Programme under Marie Sklodowska-Curie grant agreement
no. 690893.\\
A. Farajzadeh and O. Ercetin are with the Faculty of Engineering and Natural Sciences, Sabanci
University, 34956 Istanbul, Turkey (e-mail: aminfarajzadeh@sabanciuniv.edu; oercetin@sabanciuniv.edu).\\
H. Yanikomeroglu is with the Department of Systems
and Computer Engineering, Carleton University, Ottawa, ON K1S 5B6,
Canada (e-mail: halim@sce.carleton.ca).
}}

\maketitle
\begin{abstract}
Future intelligent systems will consist of a massive number of battery-less sensors, where quick and accurate aggregation of sensor data will be of paramount importance. Over-the-air computation (AirComp) is a promising technology wherein sensors concurrently transmit their measurements over the wireless channel, and a {\em reader} receives the noisy version of a function of measurements due to the superposition property. A key challenge in AirComp is the accurate power alignment of individual transmissions, addressed previously by using conventional precoding methods. In this paper, we investigate a UAV-enabled {\em backscatter} communication framework, wherein UAV acts both as a power emitter and reader.  The mobility of the reader is leveraged to replace the complicated precoding at sensors, where UAV first collects sum channel gains in the first flyover, and then, use these to estimate the actual aggregated sensor data in the second flyover. Our results demonstrate improvements of up to $10$ dB in MSE compared to that of a benchmark case where UAV is incognizant of sum channel gains.
\end{abstract}
\section{Introduction}
\lettrine[]{M}{ACHINE} type communication (MTC) is one of the disruptive technologies promised by 5G and beyond networks. MTC transmissions are uplink-dominant and
usually have low data rates. Envisioning future internet of things (IoT) networks where a massive number of sensors haphazardly deployed, will require novel approaches to quickly collect the sensed data. In most practical sensor applications (e.g., environmental monitoring), individual sensor measurements are not much of an interest but an aggregated function of them is~\cite{1632656}. Recently, over-the-air computation (AirComp) paradigm is developed as a promising approach to support fast data aggregation~\cite{8371243}. AirComp can compute a class of nomographic functions (e.g., arithmetic mean, weighted sum, geometric mean and
Euclidean norm) by coherent and simultaneous transmissions of sensors, and by exploiting the superposition property of the wireless channel~\cite{8708985}. 

Meanwhile, employing low-power and self-sustainable sensor devices are critical to support massive IoT applications. Ambient backscatter communication technology is a promising candidate for self-sustainable wireless communication systems in which there is no external power supply. By utilizing an existing radio frequency (RF) signal, this technology can support low-power sensor-type devices~\cite{8368232}.

Unmanned aerial vehicles (UAVs) have the potential to provide wireless connectivity to remote locations, and thus, they received increasing attention from the research community as well as the industry~\cite{8807380}. In the context of backscatter sensor networks, UAVs can act not only as data aggregators but also RF signal emitters enabling fast data aggregation and energy transfer at remote locations~\cite{8761125}. 
In this work, we aim to take advantage of the mobility of the UAVs to improve data aggregation from a large scale backscatter sensor network through a novel implementation of AirComp. Note that a key challenge in AirComp is the accurate power alignment to alleviate the detrimental effects of the channel attenuation. This is performed by precoding at the transmitters; however, optimal precoding requires accurate estimation of the channel gains between the sensor nodes and
the UAV. In a network with a large number of {\em backscatter} nodes, the implementation of conventional channel sensing methods such as pilot transmissions and channel estimation, are difficult if not impossible. 

Hence, in this work, we leverage the mobility of UAV in order to sample the sum channel gains from a number of different locations in a specified region of interest. Specifically, a {\em sample-then-map} mechanism is proposed, wherein UAV performs two fly-overs. In the first round of fly-over, UAV transmits a reference RF signal and backscatter nodes reflect it unaltered to the UAV. This is repeated at a number of different locations in the region of interest.  These measurements provide the UAV sum bidirectional channel gains at these locations. In the second round, the UAV transmits again a reference RF signal, but this time backscatter nodes return their actual sensor measurements.  The sensors are oblivious to the channel state and transmit sensor data without precoding.  The UAV uses  sum gains obtained in the first round to linearly combine sensor measurements in the second round by minimizing the mean square error (MSE) of the aggregated sensor data. Our results demonstrate that under realistic channel conditions, with a network of 20 sensor nodes, MSE of the proposed scheme is below -2dB, when UAV samples the network at over 4 different locations.

To the best of our knowledge, this is the first attempt to utilize UAV mobility as an enabler to improve AirComp performance without requiring computation and communication expensive precoding in backscatter sensor networks.  In ~\cite{8708985,8468002}, the authors aim at developing multiple-input-multiple-output (MIMO)
AirComp such that the objective is to find the optimal beamforming design for compensating the nonuniform fading. More recently, in~\cite{8644497}, a multi-antenna UAV-enabled AirComp is studied where UAV acts both as a data collector and wireless power transmitter. The objective was to jointly design an optimal power allocation,
energy beamforming and AirComp equalization to minimize the MSE. However, the mobility of the UAV was not taken into account in improving the MSE performance.
\begin{figure}[!t]
\centerline{\includegraphics[width=73mm,height=42.3mm,scale=1]{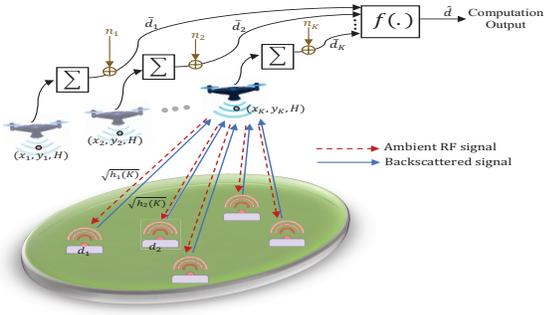}}
\caption{System Model.}
\label{sysmodel}
\end{figure}

\section{System Model}
We consider a wireless sensor network with $N$ backscatter devices distributed independently and uniformly randomly over a circular target area with radius $R_{cov}$ as given in Fig.~\ref{sysmodel}. Each node is equipped with a single sensor measuring an environmental parameter such as temperature, humidity, atmospheric pressure, etc. The node is equipped with an RF antenna receiving RF signal, and then, emitting a modulated backscatter signal. The UAV has a collocated bi-static reader and acts both as a data collector and a carrier emitter. UAV employs two separate antennas for transmission and reception operating at different frequency bands to avoid self-interference. UAV follows a given and fixed two-dimensional (2D) flight path at an altitude of $H$ meters, and a finite number, $K$, of stop-over positions, $\{(x_k,y_k)\}_{k=1}^K$, over each of which it hovers for a finite duration of time.  

In a basic backscatter channel, there are two links:  {\em Forward} (power-up) link from the UAV to a sensor node, and {\em backscatter} link from the sensor to the UAV. Most air-to-ground channel measurements and statistical models focus on large-scale statistics such as mean path-loss~\cite{8411465}. In this work, we also assume that there is no obstruction between the ground sensors and the UAV, and thus, the channels between the sensors and the UAV are assumed to be independent and identically distributed (iid) free-space path-loss channels. 

At stop-over location, $k$, UAV broadcasts a carrier signal $S(t)$, with power $P$ over the forward channel, i.e., $S(t)=\mathfrak{Re}\{\sqrt{P}e^{j(2\pi f t)}\}$ where $f$ is the carrier frequency. Let $B(t)$ be the received signal at sensor $i$, i.e., $B_i(t)=\mathfrak{Re}\{\frac{g_0\sqrt{P}e^{j(2\pi ft)}}{D_i(k)}+n(t)\}$, where $g_0$ is the channel gain at a reference distance $1$ m \cite{8379427} and $D_i(k)=\sqrt{H^2+(x_k-x_i)^2+(y_k-y_i)^2}$ is the distance between the sensor $i$ with $(x_i,y_i)$ as its coordinate, and UAV when it is at location $k$. Moreover, $n(t)$ is the additive white noise. The received power at sensor $i$ is $P_{B_i}^k=\frac{g_0^2P}{D_i^2(k)}$. Each sensor node reflects a portion of the receiver signal over the backscatter channel. At the UAV, the received signal from sensor $i$ is $Z_i(t)=\mathfrak{Re}\{\frac{g_0\sqrt{\zeta_i P}e^{j(2\pi f t)}}{D_i^2(k)}+n(t)\}$, where $\zeta_i$ is the backscatter reflection coefficient. The received power is $P_{Z_i}^k=\left(\frac{g_0^2\sqrt{P\zeta_i}}{D_i^2(k)}\right)^2$. In the rest of the letter, we drop the letter $t$ to avoid any confusion.

Hence, the overall channel power gain between the sensor $i$ and UAV when UAV is at stop-over location $k$, is $h_i(k)=\frac{g_0^2}{D^2_i(k)}$, where $i=1,\dots,N$.
\section{Over-the-Air Functional Computation}
\subsection{Overview}
In this work, our proposed method is suitable for a polynomial function of observations as the target nomographic function:
\vspace{-0.5cm}
\begin{align}\label{target}
    {d}^*=\sum_{i=1}^N w_id_i^{v_i},
\end{align}
where $w_{(.)}$ and $v_{(.)}$, are positive constants. 
In conventional applications of AirComp~\cite{8371243}, the coherent combination of multiple received data in the nomographic function is ensured by precoding the individual transmissions amplified with a gain that is reciprocal of instantaneous channel gain of the respective transmitter. 
Due to implementation and computation limitations of backscatter devices, we assume that sensors transmit plain signals without performing any precoding. Instead, we utilize the mobility of the UAV as a form of providing channel diversity. Specifically, our proposed AirComp method has two phases. In the first phase, UAV collects reference signals from multiple stop-over locations. Note that a reference signal collected by the UAV is the sum aggregate of backscatter reflections from all sensor nodes and provides sum channel gains at the respective locations. In the second phase, UAV visits the same locations, but this time each node backscatters their measured sensor value. The UAV is tasked to combine the measurements made in the second phase by assigning a linear coefficient to each measurement based on its sum channel gain observations from the first phase.
\vspace{-0.2cm}
\subsection{Mobility-assisted AirComp}
\subsubsection{Sampling Phase}
In the first phase, UAV takes $K$ noisy samples at a number of predefined locations $\{(x_k,y_k)\}_{k=1}^K$, with a pilot signal, i.e., all backscatter sensors simultaneously transmit a unit value. Hence, at sample location $k$, UAV receives
$    \sum_{i=1}^N\mathfrak{Re}\{\frac{g_0\sqrt{\zeta_i P}e^{j(2\pi f t)}}{D_i^2(k)}+n^\prime(t)\}$,
from which it determines the sum channel gain at location $k$ as $\sum_{i=1}^N g_i(k)+n^\prime_k$, where $g_i(k) = \sqrt{\zeta_i P} h_i(k)$, and
$n^{\prime}_k$ is a Gaussian distributed sampling noise with zero-mean and variance $\sigma_{n^{\prime}_k}^2$.

\subsubsection{Computation Phase}
In the second phase, UAV again collects $K$ samples from the same set of  locations $\{(x_k,y_k)\}_{k=1}^K$, but this time each sensor $i$ transmits its sensor observation $d_i$, $i=1,\dots,N$. Let sensor data be generated according to a Gaussian source with $d_i\sim \mathcal{N}(\mu_{d_i},\sigma_{d_i}^2)$. After simultaneous transmissions, the received signal at the UAV is $    \sum_{i=1}^N\mathfrak{Re}\{\frac{g_0\sqrt{\zeta_i d_i P}e^{j(2\pi f t)}}{D_i^2(k)}+n(t)\}$, where the aggregated sensor data measurements is given as:
\vspace{-0.1cm}
\begin{align}\label{est}
    \Bar{d}_k=\sum_{i=1}^{N}g_i(k)d_i+n_k,\: \forall k=1,\dots,K.
\end{align}
Based on the aggregated data measurements, the UAV obtains an estimated function $\hat{d}$, i.e.,
$    \hat{d}=f(\Bar{d}_1,\Bar{d}_2,\dots,\Bar{d}_K),$
where $f$ is a mapping used to compensate for the lack of channel precoding. In this work, we consider $f$ as a linear combination of observations such that
\vspace{-0.4cm}
\begin{align}\label{estimated}
    \hat{d}=\sum_{k=1}^K\beta_k\Bar{d}_k,
    \end{align}
    where $\beta_k, k=1,\dots,K$, are non-negative constants. Note that $\beta_k$ is basically the power coefficient assigned to the backscatter nodes by the UAV at sampling round $k$. Hence, at each round, the same power coefficient is aligned to the sensors. 
\vspace{-0.2cm}
\section{Problem Formulation and Solution}
We measure the computation error of our mobility assisted AirComp method by the MSE defined as
\vspace{-0.1cm}
\begin{align}
    &\mathbf{MSE}\big(\hat{d},{d}^*\big)=\mathbf{E}\big[\big(\hat{d}-d^*\big)^2\big]\nonumber\\
    &=\sum_{i=1}^N\sigma_{d_i}^2\big(\sum_{k=1}^K\beta_k\mathbf{E}[g_i(k)]\big)^2+\sigma_{d_i}^2\sum_{k=1}^K\beta_k^2\mathbf{Var}(g_i(k))\nonumber\\
    &+\mu_{d_i}^2\sum_{k=1}^K\beta_k^2\mathbf{Var}(g_i(k))-2\big(w_i\mathbf{E}[d_i^{v_i+1}]\sum_{k=1}^K\beta_k\mathbf{E}[g_i(k)]\nonumber\\
    &+\mu_{d_i}(\sum_{k=1}^K\beta_k\mathbf{E}[g_i(k)])(\sum_{j\neq i}^N w_j\mathbf{E}[d_j^{v_j}])\big)+w_i\mathbf{Var}(d_i^{v_i})\nonumber\\&+(\sum_{i=1}^Nw_i\mathbf{E}[d_i^{v_i}])^2+\sum_{k=1}^K\beta_k^2\sigma_{n_k}^2,
\end{align}
where $\mathbf{E}[.]$ and $\mathbf{Var}(.)$ are expectation and variance operators, respectively.  The expectations are over the random locations of sensor nodes and their random observations. Note that since $d_{(.)}$ is a Gaussian source, for any values of $v_i$, $i=1,\dots,N$, the corresponding moments can be calculated accordingly. Hence, considering any polynomial target function, the problem formulation is the same and the power alignment can be applied; however, for each target function, the power coefficient $\beta_k$ takes a unique and different value. For instance, for the case $v_i=2$, the target function becomes the sum of weighted squared sensor observations and the MSE can be simplified such that $\mathbf{E}[d_{(.)}^2]{=}\mu_{d_{(.)}}^2{+}\sigma_{d_{(.)}}^2$, $\mathbf{E}[d_{(.)}^3]{=}\mu_{d_{(.)}}^3{+}3\mu_{d_{(.)}}\sigma_{d_{(.)}}^2$, and $\mathbf{Var}(d_{(.)}^2){=}2\sigma_{(.)}^2(\mu_{(.)}^2{+}\sigma_{(.)}^2)$.

Our objective is to minimize MSE by choosing coefficients $\beta_k$ for a given UAV trajectory plan $\{(x_k,y_k)\}_{k=1}^K$. Thus, the optimization problem can be formulated as
\vspace{-0.2cm}
\begin{subequations}
\begin{align}\label{opt1}
&\operatorname*{min}_{\substack{\{ \beta_k\}^{K}_{k=1}}}
\mathbf{MSE}\big(\hat{d},{d}^*\big)\\
& \text{s.t.}
\sum_{k=1}^K\beta_k\leq \beta_0,\label{const2}
\end{align}
\end{subequations}
where constraint \eqref{const2} represents the limited receiver power. 
\vspace{-0.3cm}
\subsection{Special Case}
Note that the problem \eqref{opt1}-\eqref{const2} is not amenable to a closed form solution, since additional statistical information about sensor locations is required and this is usually not available \cite{8741612}. Moreover, when the power coefficients are equally aligned to backscatter devices, i.e., $\beta_1\:{=}\:\dots\:{=}\:\beta_K\:{=}\:\beta$, the problem can be shown to be convex and an optimal solution can be calculated according to Theorem~\ref{Theo1}.
\begin{thm}\label{Theo1}
 For $\beta_1\:{=}\:\dots\:{=}\:\beta_K\:{=}\:\beta$, the optimal solution of \eqref{opt1}-\eqref{const2} is
\vspace{-0.2cm}
\begin{align}\label{beta}
&\beta^*=\nonumber\\&\frac{\sum_{i=1}^N\sum_{k=1}^K\mathbf{E}[g_i(k)]\big (w_i\mathbf{E}[d_i^{v_i+1}]{+}\mu_{d_i}\sum_{j\neq i}^Nw_j\mathbf{E}[d_j^{v_j}]\big)}{\sum_{i=1}^N\sigma_{d_i}^2(\sum_{k=1}^K\mathbf{E}[g_i(k)])^2{+}\big(\sigma_{d_i}^2+\mu_{d_i}^2\big)\sum_{k=1}^K\mathbf{Var}(g_i(k))},
\end{align}
\vspace{-0.5cm}
\end{thm}
\proof See Appendix.\\
Even under this assumption, due to the involvement
of statistical operations, the MSE expression
is too complicated to draw insights. To resolve this issue, in
the following we propose a heuristic method to resort to approximate channel power gains.
\vspace{-0.6cm}
\subsection{Heuristic Approach}
Note that channel gains are based on free-space path-loss model.  The distance between the UAV and each node lays in the interval of $H{\leq} D_{(.)}{\leq} H\sqrt{1+(\frac{2R_{cov}}{H})^2}$. At high altitudes when $H\:{>}\:4R_{cov}$, it can be shown that $H{\leq} D_{(.)}{\leq} H\sqrt{1+\epsilon^2}$, $\epsilon{<}1$. Hence, the difference between the path-loss values is less than $5$\%. Due to the relatively high altitude of the UAV compared to the girth of the area of coverage, the contributions from each sensor node in the initial sampling phase can be considered approximately the same, i.e., if $\sum_{i=1}^N g_i(k)+n^\prime_k=\alpha_k$, then the channel gain of each sensor node $i$ at location $k$ can be approximated as
    $g_i(k)\approx\frac{\alpha_k}{N}$, for all $i,k$.
Consequently, with this approximation, one can show that \eqref{opt1}-\eqref{const2} is convex with respect to $\beta_k$, $k=1,\dots,K$, and use the first order Karush Kuhn Tucker conditions to obtain a sub-optimal solution as
\vspace{-0.2cm}
    \begin{align}
        &\beta_k=\nonumber\\&\frac{\sum_{i=1}^N (w_i\mathbf{E}[d_i^{v_i+1}]{+}\mu_{d_i}\sum_{j\neq i}^N w_j \mathbf{E}[d_j^{v_j}]){-}\sum_{k^\prime\neq k}^K\beta_{k^\prime}\frac{\alpha_{k^\prime}}{N}\sum_{i=1}^N\sigma_{d_i}^2}{\frac{\alpha_k}{N}\sum_{i=1}^N\sigma_{d_i}^2+\frac{N\sigma_{n_k}^2}{\alpha_k}}.
    \label{beta_heur}
    \end{align}

    This implies that at sampling round $k$ in the computation phase, power coefficient $\beta_k$ is aligned to all nodes by UAV. Note that \eqref{beta_heur} has multiple solutions. Hence, we further suppose that $\sum_{k^\prime\neq k}\beta_{k^\prime}\alpha_{k^\prime}=(K-1)\beta_k\alpha_k$, which roughly means that each individual sampling phase contribution is the same, and it leads to the following unique $\beta_k$ values.
\vspace{-0.17cm}
\begin{align}
    \beta_k=\frac{\sum_{i=1}^Nw_i\mathbf{E}[d_i^{v_i+1}]+\mu_{d_i}\sum_{j\neq i}^N w_j \mathbf{E}[d_j^{v_j}]}{\frac{\alpha_k}{N}\sum_{i=1}^N\sigma_{d_i}^2+\frac{N\sigma_{n_k}^2}{\alpha_k}+(K-1)\frac{\alpha_k}{N}\sum_{i=1}^N\sigma_{d_i}^2}.
\end{align}
Moreover, if $\beta_1=\dots=\beta_K=\beta$, then we have 
\vspace{-0.17cm}
\begin{align}
    \beta =\frac{\sum_{i=1}^Nw_i\mathbf{E}[d_i^{v_i+1}]+\mu_{d_i}\sum_{j\neq i}^N w_j \mathbf{E}[d_j^{v_j}]}{\frac{1}{N}\sum_{k=1}^K\alpha_k\sum_{i=1}^N\sigma_{d_i}^2+\frac{N\sum_{k=1}^K\sigma_{n_k}^2}{\sum_{k=1}^K\alpha_k}}.
\end{align}
\vspace{-0.5cm}
\section{Numerical Results}
In this section, we evaluate the MSE performance with respect
to the number of samples $K$ in which the UAV takes, considering three different target functions: 1) The sum of observations, i.e., $w_i\:{=}\:1,\:v_i\:{=}\:1$; 2) The sum of equally-weighted squared observations, i.e., $w_i\:{=}\:1,\:v_i\:{=}\:2$; 3) The sum of unequally-weighted cubed observations, i.e., $w_i\in\{1,2,\dots,N\}$, $v_i\:{=}\:3$. Moreover, we also examine the MSE performance with respect to the number of backscatter sensors $N$ served by the UAV. We assume that UAV moves within the field of interest on a fixed and predefined 2D path along the diameter of the field. The network parameters considered in these experiments are $P\:{=}\:30$ dBm, $\sigma_{n}^2\:{=}\:-70$ dBm, $H\:{=}\:50$ m, $R_{cov}\:{=}\:10$ m, $K\:{=}\:5$, $N\:{=}\:20$, and $\zeta_{(.)}\:{=}\:0.99$. Moreover, since backscatter systems mainly operate at $f\:{=}\:868$ MHz, we consider $g_0\:{=}\:0.0275$ ($g_0\:{=}\:\frac{c}{4\pi f}$). In Fig.~\ref{fig1}, the MSE performance is plotted with respect to $N$. For comparison purposes, we use a mapping function that takes average of all sensor measurements without weighing them with respect to their locations. Fig.~\ref{fig1} illustrates that when the linear mapping coefficients $\beta_k$, $k=1,\dots,K,$ are not chosen carefully, i.e., $\beta_1\:{=}\:\dots\:{=}\:\beta_K$, due to loss of power coefficient alignment, there is around $10$ and $1$ dB reduction in the MSE value compared to the benchmark case and linear weighted mapping case, respectively. Moreover, we observe that increasing the number of sampling locations do not always improve the MSE for target functions when it is increased beyond a certain point. It is also observed that the MSE decreases when higher order polynomial functions are considered as target function due to the degraded performance of proposed linear mapping function in compensating the channel. For example, when the sum of unequally-weighted cubed observation is considered, there is up to $5$ dB reduction in MSE compared to the case that the sum of unity-weighted observations is considered.\\
\indent In Fig.~\ref{fig2}, we examine the performance of the MSE with respect to the number of backscatter sensors $K$ considering three different target functions as aforementioned. It can be seen that as the number of backscatter sensors increases, there is always more than $5$ dB difference in MSE performance over the benchmark case. Moreover, the rate of increase in MSE is almost the same for both of cases where linear mapping coefficients are considered to be equal or chosen individually.  
\vspace{-0.5cm}
\section{Conclusion}
In this letter, we introduced the idea of using mobility as a diversity mechanism to improve the performance of AirComp systems when channel gains cannot be measured. Our approach is especially suitable for large-scale backscatter sensor networks served by a UAV flying over a predetermined path. Our numerical results demonstrate that our proposed mobility assisted AirComp improves over conventional AirComp significantly in terms of target MSE. As a future work, we will consider joint optimization of UAV location, backscattering reflection coefficient selection, and linear sampling coefficient selection policies. Another direction is to aim to find an optimal channel-inversion function $f(\cdot)$ such that MSE is minimized.
\vspace{-0.3cm}
\appendix
\label{app1a}
Note that $\frac{\partial^2\textrm{MSE}(\hat{d},d^*)}{\partial \beta^2}\geq 0$, hence, the problem~\eqref{opt1}-\eqref{const2} is a convex problem with respect to $\beta$ and can be solved by using Lagrangian method. Let $\mathcal{L}(\beta,\gamma)$ be the Lagrangian function expressed as 
     $\mathcal{L}(\beta,\gamma)=\sum_{i=1}^N\sigma_{d_i}^2\beta^2(\sum_{k=1}^K\mathbf{E}[g_i(k)])^2+\sigma_{d_i}^2\beta^2\sum_{k=1}^K\mathbf{Var}(g_i(k))-2\big(\beta w_i\mathbf{E}[d_i^{v_i}]\sum_{k=1}^K\mathbf{E}[g_i(k)]+\mu_{d_i}\beta\sum_{k=1}^K\mathbf{E}[g_i(k)]\sum_{j\neq i} w_j\mathbf{E}[d_j^{v_j}]\big)+w_i\mathbf{Var}(d_i^{v_i})+(\sum_{i=1}^Nw_i\mathbf{E}[d_i^{v_i}])^2{+}\beta^2\sum_{k=1}^K\sigma_{n_k}^2{-}\gamma(K\beta-\beta_0),$
where $\gamma\geq 0$ is the Lagrangian multiplier. Following the KKT conditions, i.e., $\frac{\partial \mathcal{L}(\beta,\gamma)}
    {\partial \beta}=0, \gamma(K\beta-\beta_0)=0$,
the optimal solution can be determined as Eq.~\eqref{beta} when $\gamma=0$.
\begin{figure}[!t]
\centerline{\includegraphics[width=95mm,scale=10]{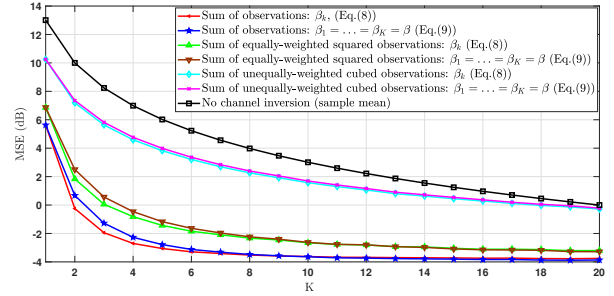}}
\caption{MSE performance vs. the number of samples
$K$ ($N=20$).}
\label{fig1}
\end{figure}
\begin{figure}[!t]
\centerline{\includegraphics[width=95mm,scale=1]{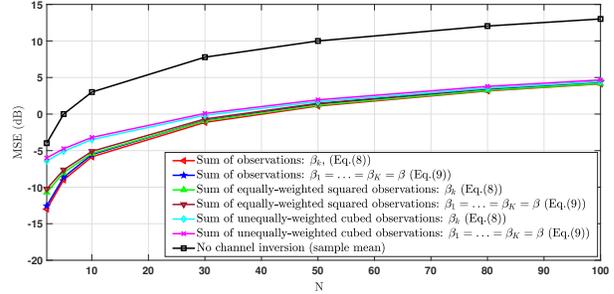}}
\caption{MSE performance vs. the number of sensors $N$ ($K=5$).}
\label{fig2}
\end{figure}
\vspace{-0.07cm}
\bibliographystyle{IEEEtran}
\bibliography{refer}
\end{document}